\documentclass[review]{elsarticle}

\usepackage{lineno,hyperref}
\modulolinenumbers[5]
\usepackage{graphicx}
\usepackage{epstopdf}
\usepackage{amsmath}
\usepackage{rotating}
\usepackage{color}

\journal{Journal of \LaTeX\ Templates}

\begin{document}

\begin{frontmatter}

\title{Foams Stabilized by Tricationic Amphiphilic Surfactants\tnoteref{mytitlenote}}
\tnotetext[mytitlenote]{Fully documented templates are available in the elsarticle package on \href{http://www.ctan.org/tex-archive/macros/latex/contrib/elsarticle}{CTAN}.}


\author[physicsdept]{Seth Heerschap}
\author[chemistrydept]{John N. Marafino}
\author[chemistrydept]{Kristin McKenna}
\author[chemistrydept]{Kevin L. Caran}

\author[physicsdept]{Klebert Feitosa
\corref{mycorrespondingauthor}}
\cortext[mycorrespondingauthor]{Corresponding author}
\ead{feitoskb@jmu.edu}


\address[physicsdept]{Dept. of Physics and Astronomy, James Madison University, 901 Carrier Dr., MSC 4502, Harrisonburg, VA 22807}

\address[chemistrydept]{Dept. of Chemistry \& Biochemistry, James Madison University, 901 Carrier Dr., 
MSC 4501,
Harrisonburg, VA 22807}

\begin{abstract}
The unique surface properties of amphiphilic molecules have made them widely used in applications where foaming, emulsifying or coating processes are needed. Novel surfactant architectures with multi-cephalic and multi-tailed molecules have reportedly enhanced their anti-bacterial activity in connection with tail length and the nature of the head group, but their ability to produce and stabilize foam is mostly unknown. Here we report on experiments with tris-cationic, triple-headed, double- and single-tailed amphiphiles and their foamability and foam stability with respect to head group, tail number and tail length. The amphiphiles are composed of an aromatic mesitylene core and three benzylic amonium bromide groups, with alkyl chains attached to one or two of the head groups.   
Whereas shorter (14 carbons in length) double-tailed molecules are found to produce very stable foams, foams made with single tail molecules of the same length show poor foamability and stability, and foams with longer (16 carbons in length) double-tail molecules do not foam with the methods used. By contrast, the structure of the non-tail-bearing head group (trimethylammonium vs. pyridinium) has no impact on foamability. Furthermore, observations of the coarsening rate at nearly constant liquid content indicate that the enhanced foam stability is a result of lower gas permeability through the surfactant monolayer. Finally, the critical aggregation concentration (CAC) of the surfactants demonstrates to be a good predictor of foamability and foam stability for these small molecule surfactants. This results inform how surfactant architecture can be tailored to produce stable foams.
\end{abstract}

\begin{keyword}
\texttt{elsarticle.cls}\sep Foams \sep Surfactant monolayers \sep Stability
\MSC[2010] 00-01\sep  99-00
\end{keyword}

\end{frontmatter}


\section{Introduction}

The architecture of an amphiphile profoundly affects its properties. Molecules with multiple polar headgroups and/or non-polar tails have different, and sometimes unexpected, aggregation characteristics relative to “conventional” amphiphiles with one head and one tail. For example, gemini amphiphiles (composed of two conventional amphiphiles connected via a covalent linker between the two heads) generally form aggregates in aqueous solution at significantly lower concentrations than their conventional counterpart \cite{Menger2000Gemini,Shukla2006Cationic}.  Recently, we \cite{Marafino2015Colloidal, LaDow2011Bicephalic},  and others \cite{Bhattacharya2011Surfactants,
Haldar2005Synthesis,Muckom2013Dendritic,
Maisuria2011Comparing,Sugandhi2007Synthesis,
Paniak2014Antimicrobial}  have been interested in developing an understanding of the structure/property relationships of amphiphiles with various numbers of heads and tails. Modifying the architecture of an amphiphile affects its ability to align at an interface, so we predicted that using compounds with non-conventional structures would allow us to tailor the properties of aqueous foams. The work presented here includes studies on the dynamics of foams of conventional amphiphiles, as well as those with three cationic headgroups and one or two non-polar tails. 

Liquid foams are a dispersion of gas bubbles in aqueous solutions stabilized by surface-active molecules \cite{Weaire2001Physics}.
Foams have important application in diverse industries raging from consumer products (detergents, cosmetics, food), to metalurgy and mining \cite{Cantat2013Foams,2012Foam}. Many of these applications requires a stable foam which depends on the physical-chemical properties of the surfactants. Due to their amphiphilic nature surfactants adsorb at the air-water interfaces and stabilize the foam films by exerting a disjointing pressure that counterbalances van der Walls attractive forces \cite{Langevin2000Influence}. The concentration and adsorption energy of surfactants determines how easily the foam is produced. Once the foam is made, three physical processes control its subsequent evolution and final fate: liquid drainage, coarsening and coalescence. These processes are not independent from the chemical properties of the surfactants, but the connection is not yet firmly established. Generally the drainage of liquid along the the network of intersecting films (called ``Plateau borders'' \cite{Addad2013Flow}) depends not only on the viscosity of the fluids, but also on the interfacial mobility of the surfactant molecules \cite{Addad2013Flow,Stone2003Perspectives}. Foam coarsening is a result of gas diffusion driven by Laplace pressure differences between bubbles. The rate of bubble growth is primarily a function of the gas diffusivity and solubility in the liquid, and, to a lesser extent, the gas permeability of the surfactant monolayer \cite{Tcholakova2011Control,Rio2014Unusually}. Coarsening is also strongly coupled to, and works in cooperation with, drainage to accelerate foam aging \cite{Hilgenfeldt2001Dynamics,Vera2002Enhanced,Feitosa2008Gas,SaintJalmes2006Physical}.  
As the bubbles grow and liquid drains, the films become unstable and rupture leading to bubble coalescence and the gradual disappearance of the foam. This process has been attributed to the desorption energy of the surfactant at the film interfaces \cite{Langevin2000Influence,Langevin2015Bubble}. A high desorption energy induces high interfacial viscoelasticity, which inhibits coarsening, traps liquid in the films and makes the foam more stable.

Many studies of foam evolution have not considered the contribution of surfactant molecules to foam stability, mainly because drainage and coarsening depend more strongly on the physical properties of the foam---the average bubble size and liquid fraction---than on the chemical properties of surfactants \cite{SaintJalmes2006Physical}. Recent studies, however, have demonstrated that the molecular structure of surfactants can have a non-negligible impact on foam evolution and stability \cite{Martin2002Network,Bhattacharyya2000Surface,Davis2007Comparisons,Croguennec2006Interfacial}. In particular, a systematic study of foams made with a series of oligomers of the surfactant DTAB reported by Salonen and co-workers \cite{Salonen2010Solutions} shows a clear connection of foamability and foam stability with surfactant structure, supporting the idea that surfactant architecture can even be designed to tune foam stability.

We report on the foamability and foam stability of a new class of multi-cephalic and multi-tailed surfactants to determine how the complex architecture of these surfactants impact foam stability. Over the course of the experiment we track changes in surface liquid fraction, foam volume, bubble size and distribution as a function of time. The results show that while the identity of surfactant's head group has no noticeable impact on foam properties, the number of tails and tail length are the major features affecting the foamability and stability of the foam. A careful analysis of the coarsening data indicates that the stability is connected to a relatively low gas diffusion rate between the bubbles driven by the low gas permeability of the surfactant monolayer, in contrast with results reported with conventional surfactants where the gas transport across the interface plays a minor role on foam stability. Moreover, the correlation between the critical aggregation concentration (CAC) and foam properties supports early evidence that,
for small molecule surfactants, the CAC could be a good predictor of foambility and foam stability.

\section{Materials and method}

\paragraph{Surfactant compounds} The amphiphiles used in this study consist of three cationic benzylic amonium bromide head groups connected to a mesitylene core. The compounds are named as \textit{M-X,n,n}, where \textit{M} refers to the mesitylene core, and \textit{X} and \textit{n} to the attached groups. One of the head groups is either pyridinium ($X=P$) or trimethylammonium ($X=1$). The others can be a trimethylammonium group ($n=1$), or a dimethylalkylammonium group with $n$ (14 or 16) representing the number of carbon atoms in the long linear hydrocarbon tail. Different combinations of head group, tail length and number of tails were explored using four different surfactants: (M-1,14,14), (M-P,14,14), (M-1,16,16) and (M-1,1,14). Details of their synthesis are described in reference~\cite{Marafino2015Colloidal,Caran2015Inprep}. The foam properties of the surfactants above are compared with foams made with commercially available sodium dodecylsulfate (SDS), an anionic molecule, and cetyltrimethylammonium bromide (CTAB) and dodecyltrimethylammonium bromide (DTAB), two molecules that, similar to the triple headed amphiphiles, have quaternaryammonium headgroups, bromide counterions and tails of comparable length. All the molecules were dissolved in de-ionized water. The molecular mass, critical aggregation concentration (CAC) and molecular structure of the surfactants are given in Table \ref{tab:surfactants}. Initial investigations of the stability of SDS foams produced by vigorous handshake showed that 0.100 M solutions produced stable foams lasting for several hours. This concentration corresponds to about 12.2 times the CAC of SDS. To effectively compare stability across surfactants, all solutions were prepared at 12.2$\times$CAC. All experiments were performed at approximately 22$^\circ$C.

\begin{sidewaystable}
\begin{tabular}{|p{52pt}|p{30pt}|p{35pt}|p{70pt}|p{35pt}|p{110pt}|p{50pt}|p{40pt}|}
\hline
Molecule &  Head charge & Tail length & Molecular mass [g$\cdot $mol$^{-1}$] & CAC [mM] & Molecular structure & Foamability & Stability\\
\hline
\hline
M-1,14,14 & +3 & (2) 14  & 898.90 & 0.61$^{(a)}$ & \includegraphics[width=.2\textwidth]{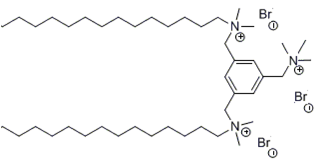} & Good & Good \\
\hline
M-P,14,14 & +3 & (2) 14 & 918.89 & 0.60$^{(a)}$ & \includegraphics[width=.2\textwidth]{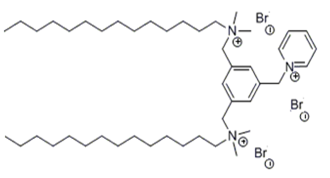} & Good & Good \\
\hline
M-1,16,16 & +3 & (2) 16 & 955.01 & 0.16$^{(a)}$ & \includegraphics[width=.2\textwidth]{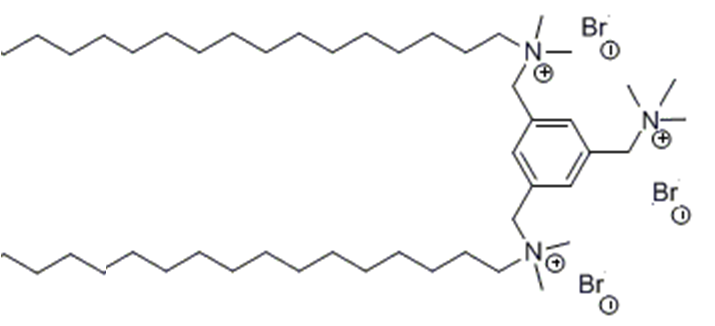} & No foam & ---\\
\hline
M-1,1,14 & +3 & (1) 14 & 716.56 & 21$^{(a)}$ &  \includegraphics[width=.2\textwidth]{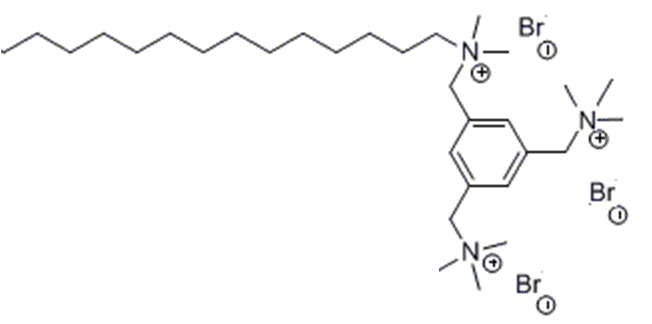} & Poor & Poor \\
\hline
\hline
SDS & -1 & (1) 12 & 288.38 & 8.2$^{(b)}$ & \includegraphics[width=.2\textwidth]{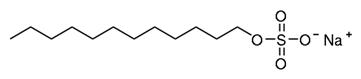} & Good & Medium\\
\hline
CTAB & +1 & (1) 16 & 364.46 & 0.98$^{(b)}$ & \includegraphics[width=.2\textwidth]{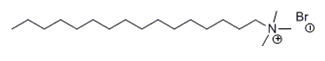} & Good & Good \\
\hline
DTAB & +1 & (1) 12 & 308.34 & 16$^{(b)}$ & \includegraphics[width=.2\textwidth]{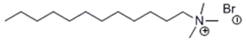} & Good & Marginal\\
\hline
\end{tabular}
\caption{Summary of properties of molecules. The number in parenthesis indicates the number of tails. CAC measured at (a) 37$^\circ$C \cite{Marafino2015Colloidal,Caran2015Inprep} and (b) 25$^\circ$C \cite{Rosen2004Surfactants}.}
\label{tab:surfactants}
\end{sidewaystable}

\begin{figure}
\begin{center}
\includegraphics[width=.8\textwidth]{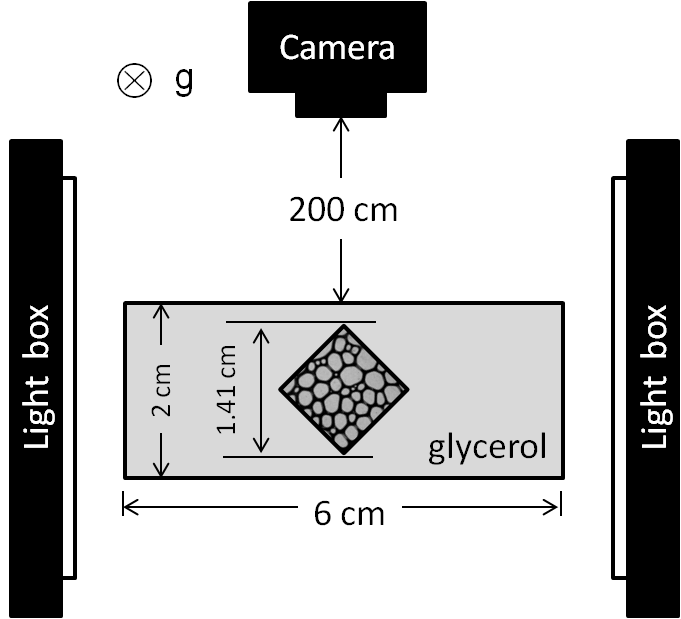}
\caption{Schematic diagram of a top-down view of the experimental set up (not to scale). The square container in the center is a cuvette cell filled with foam immersed in a tank with glycerol. Light boxes placed on each side of the tank a short distance away provide uniform illumination. Images of the foam are taken with a camera positioned directly across from the tank.}
\label{fig:cartoon}
\end{center}
\end{figure}

\paragraph{Experimental set up} Foam evolution is strongly dependent on initial conditions. In order to  better
control the initial bubble size distribution and foam liquid content, we used standard microfluidic techniques \cite{Garstecki2004Formation} to produce the foam
for all but the M-1,1,14 solution, which foams only by vigorous shaking, and the M-1,16,16 solution, which does not foam at all.
The microfluidic device consisted of a cross channel, 710 $\mu$m wide, etched (VersaLaser printer) on a PMMA substrate. The surfactant solution was pumped (Harvard PHD 2000) on opposite inlets of the cross channel at a rate of 5.91 mL/h, while ambient air was pumped (New Era Pumping Systems NE-4000) onto the adjacent inlet at a rate of 76.16 mL/h. A clear square cuvette of internal dimensions 10.6 mm $\times$ 10.6 mm $\times$ 40.0 mm was filled with the foam collected at the outlet and sealed and then immersed in a tank with glycerol (99\%, Fisher Scientific).
This marks the initial time.
As illustrated in Fig.~\ref{fig:cartoon}, two light boxes (Vista VT-12A and a Dolan Jenner QVABL connected to a Dicon LED source) placed at opposite sides of the tank (25 cm and 7 cm away) provided steady, uniform illumination. A Nikkon D-200 camera with a AF-S Micro Nikkor 105 mm 1:2.8G ED lens was pointed at the face of the tank, 90$^\circ$ from the light box. Pictures of the foam were taken every two minutes for up to 400 minutes depending on how stable the foam was. 

\paragraph{Foam imaging} In a container filed with foam, pictures of surface bubbles are marred by bubbles in adjacent internal layers which interfere with their imaging. To obtain clear images of the surface bubbles we use a technique introduced by Roth et al. \cite{Roth2013Structure} based on total internal reflection (TIR). Due to the low critical angle for TIR between glycerol and air (42.6$^\circ$), a light ray striking an air bubble will specularly reflect towards the camera, whereas a light ray striking the surface of a water filled Plateau border will be diverted into another direction. By positioning the cuvette cell 45$^\circ$ away from the camera we were able to perform simultaneous visualization of two of its four sides.

\begin{figure}
\begin{center}
\includegraphics[width=.8\textwidth]{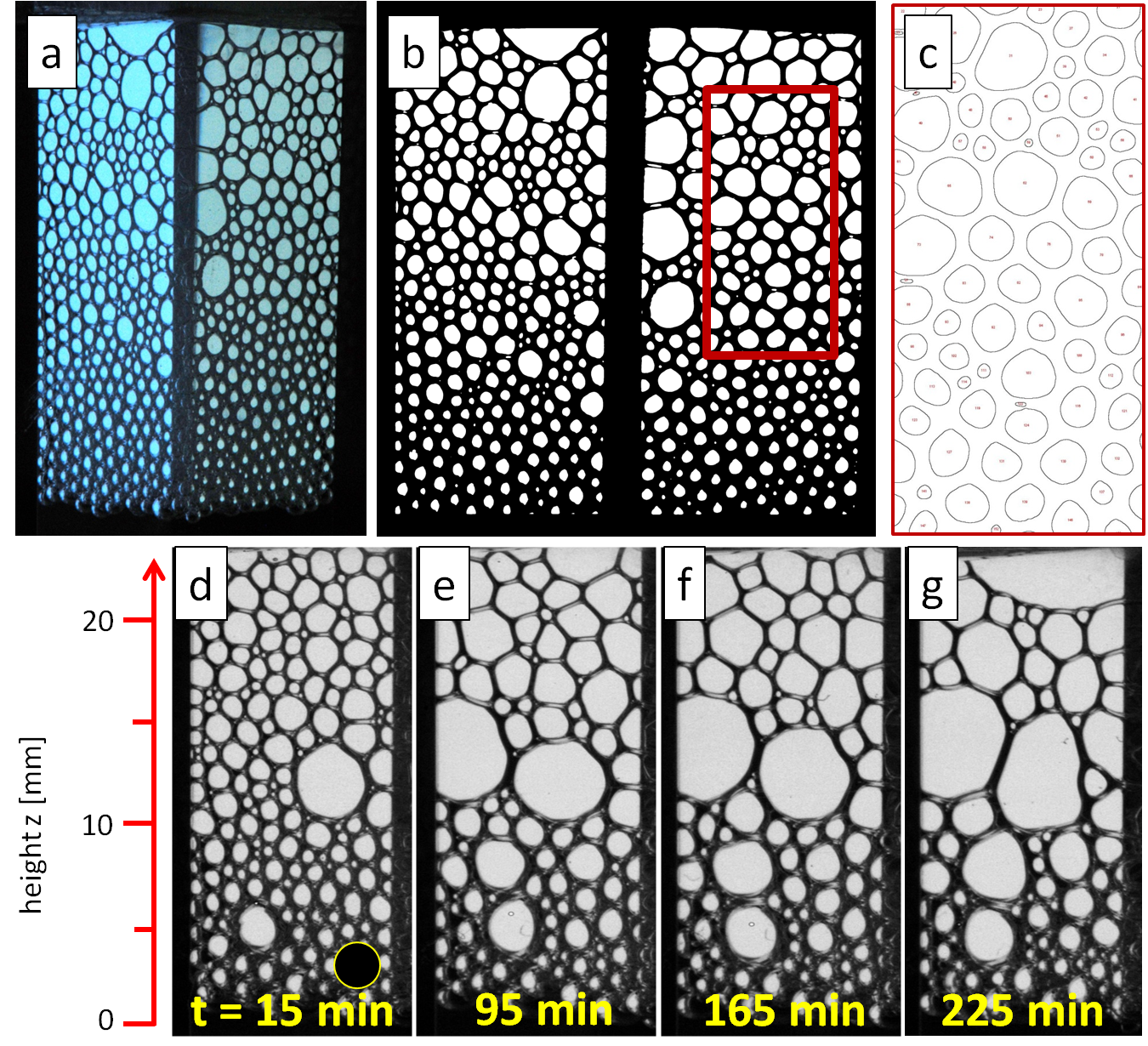}
\caption{(a) Raw image of a 15 min old SDS foam; (b) binary image corrected for angular distortion; (c) outlines of identified bubbles in the region corresponding to the red box on the binary image. (d-g) Series of images of the M-1,14,14 foam illustrating the coarsening process and collapse of the top layers. The black circle in (d) represents the average bubble size $R_{s}$.}
\label{fig:image_analysis}
\end{center}
\end{figure}

\paragraph{Image processing} We used the public domain image processing software \textit{ImageJ} (1.47u) to analyse the images. A raw image of the foam is shown in Fig. \ref{fig:image_analysis}a. Note that the image appears distorted due to the angle of the cuvette. Prior to analysis, the  images were converted to grayscale, thresholded and corrected for angular distortion. Figure \ref{fig:image_analysis}b shows the resulting processed image. Bubble surface areas were then obtained by performing the operation \textit{Analyse Particles} in ImageJ which measures area, centroid, circumference, and circularity of disconnected objects in the binary images. Objects with circularity~$<0.4$ and area~$<300$ pixels ($\sim 0.007$ mm$^2$) were excluded. Outlines of identified bubbles are shown in Fig. \ref{fig:image_analysis}c. The equivalent radius $R_i$ was then obtained from the computed areas $A_i$ as $R_i=\sqrt{A_i/\pi}$. In experimental studies where surface bubbles of 3-dimensional poly-disperse foams are imaged, it is customary to use the Sauter mean radius $R_s=\langle R^3\rangle/\langle R^2 \rangle$ to represent the mean bubble size \cite{Feitosa2008Gas}. The black circle in Fig. \ref{fig:image_analysis}d shows $R_{s}$ obtained for the bubbles shown in the figure. 
The initial average bubble radius for the various samples with the microfluidics apparatus was 0.64$\pm$0.19 mm. Note that the variability in size is caused by small variations on pumping rate and bubble coarsening during the cell filling time (approx. 2 min). The initial average bubble radius for the sample produced by shaking (M-1,1,14) was 0.38$\pm$0.12 mm.
Careful visual inspection of the foam revealed that bubbles experience coarsening in the bulk and rupture in the top layers as illustrated in the series of snapshots of the foam shown in Fig. \ref{fig:image_analysis}d-g. Bubbles also occasionally come into view or disappear into the bulk as a result of bubble rearrangements. These rearrangements, however, do not change 
the average bubble size or the surface covered by the foam, defined as $S=\sum A_i$.

\begin{figure}[h]
\begin{center}
\includegraphics[width=.8\textwidth]{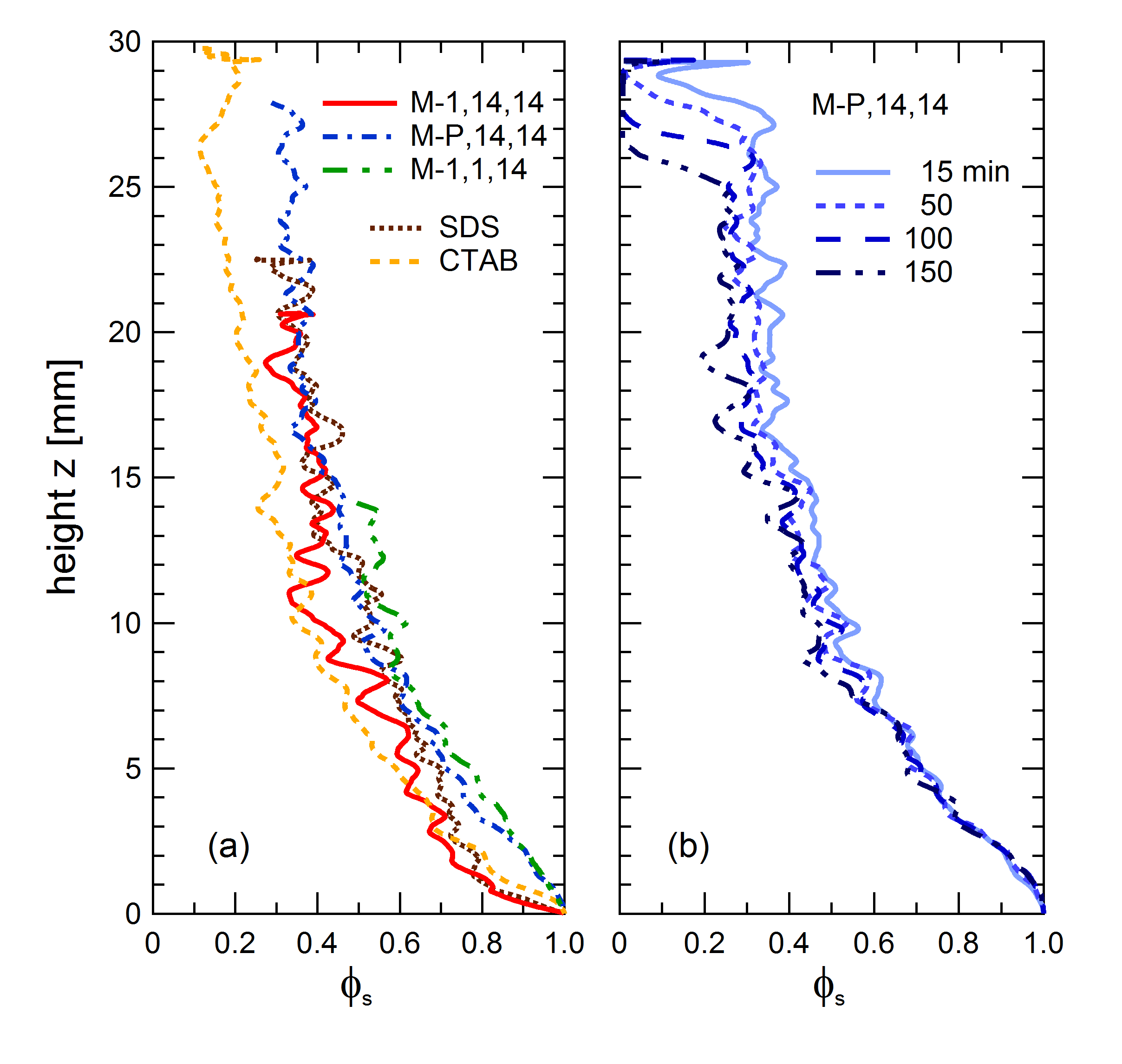}
\caption{(a) Vertical profile of the surface liquid fraction for the various foams at $t=15$~min; (b) vertical profile of the surface liquid fraction at various times for the M-P,14,14 foam; $z=0$ is defined as the foam/bulk liquid interface.}
\label{Fig:liquid_fraction_all}
\end{center}
\end{figure}

\paragraph{Liquid fraction} Traditional methods of measuring liquid fraction of foams based on  conductivity \cite{Feitosa2005Electrical} proved to be difficult to implement in the small cuvette cell. Thus, we measured the surface liquid fraction profile, $\phi_s(z)$, by computing the percentage coverage area of Plateau borders. While an explicit relationship between the surface liquid coverage and the bulk liquid fraction in 3D foams has been proposed for dry foams \cite{Weaire2001Physics}, an equivalent function for wet foams is much more difficult to establish. Nevertheless, the surface liquid fraction allows us to compare the relative wetness of the different solutions and track its evolution in time. In Fig. \ref{Fig:liquid_fraction_all}a, $\phi_s(z)$ is plotted for all solutions 15 min after the foams have been made,
when all foams measurements become synchronous.
To reduce noise, the curves were smoothed via a 300-point unweighted sliding average, where each point was replaced by the average of its neighbors within the window. Note that the wetness decreases with height as a result of drainage. In addition, the solutions show similar profiles with minor fluctuations due to variations in bubble size. We also monitored changes in the surface liquid fraction as a function of time. Figure~\ref{Fig:liquid_fraction_all}b shows a typical result here obtained for the M-P,14,14 foam. Over the course of the experiment, the surface liquid fraction profile changes only slightly due to both drainage and coarsening. The abrupt drop in $\phi_s$ in the top layer is a result of bubble collapse.
Initial measurements of the foams covered by this study, including average liquid fraction, are presented in Table \ref{tab:initial_cond}.

\begin{table}
\begin{center}
\begin{tabular}{lcccc}
\hline
Molecule & $\overline{\phi}_s$ & $R_{s0}$ [mm] & $p$ & foam age [min]\\
\hline
\hline
M-1,14,14 & 0.49 & 0.66 & 0.25 & 8\\
\hline
M-P,14,14 & 0.52 & 0.60 & 0.14 & 6\\
\hline
M-1,1,14 & 0.71 & 0.38 & 0.37 & 1\\
\hline
SDS & 0.58 & 0.41 & 0.22 & 4\\
\hline
CTAB & 0.40 & 0.87 & 0.22 & 4\\
\hline
\end{tabular}
\caption{First measurement of average liquid fraction, average radius and polydispersivity of the different foams taken at the corresponding foam age computed from the time the cuvette cell was sealed. Note that the M-1,1,14 foam was produced by shaking whereas the others by microfluidics.}
\label{tab:initial_cond}
\end{center}
\end{table}

\section{Results}

First, we qualitatively investigated the foambility of the different surfactants. Solutions of M-1,14,14 and M-P,14,14 produced a consistent  uniform foam with the microfluidic apparatus, as did solutions with SDS, CTAB and DTAB. These were thus classified as having good foamability. Solutions of M-1,1,14 did not produce foam in the microfluidic apparatus, but foamed if the cuvette cell was vigorously shaken. So, its foamability was classified as poor. Solutions made with the surfactant M-1,16,16 did not foam with the microfluidic device and hardly foamed at all by vigorously shaking. Due to its extremely poor foamability this surfactant was excluded from subsequent studies. These results are summarized in Table~\ref{tab:surfactants}.

\begin{figure}
\begin{center}
\includegraphics[width=1\textwidth]{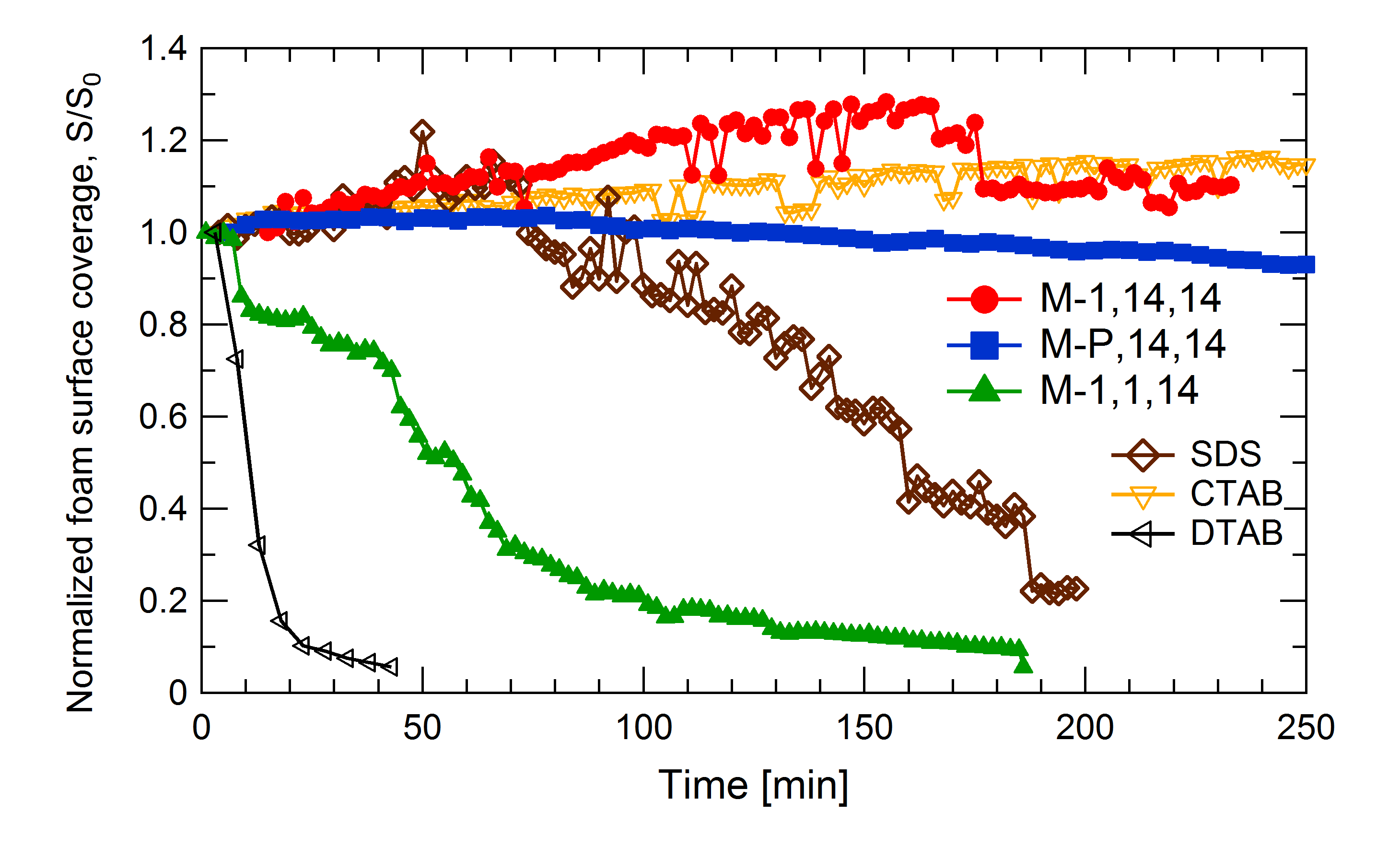}
\caption{Normalized foam covered surface as a function of time for various foams.}
\label{fig:foaming_parameter}
\end{center}
\end{figure}

Having determined the solutions that produced good foam, we turned our attention to stability. Foam stability is determined by the rate of film rupture, in particular on the top layers. This process leads to a decrease in foam volume and consequently the percentage of wall surface covered by the foam over time. Thus, we track foam stability by measuring the normalized foam surface coverage, $S/S_0$, where $S_0$ is the initial foam coverage. Figure \ref{fig:foaming_parameter} shows $S/S_0$ for the different foams as a function of time. We observe that the M-1,14,14, M-P,14,14 and CTAB foams have good stability, lasting longer than four hours. By comparison, SDS foam has only moderate stability, and starts to decay steadily after about 90 minutes. The M-1,1,14 foam shows much poorer stability and dissipates to less than 20\% of its initial coverage in about 90 minutes. Lastly, DTAB foam is by far the least stable of all, vanishing almost completely shortly after its production. Note that the apparent increase in surface coverage with time observed in the stable foams is due to the finite thickness of the Plateau borders.
Recall that
the area coverage is defined as the sum of the areas of individual bubbles. As time passes, liquid drains, plateau borders get thinner, bubbles get bigger and the total area covered by the bubble films gets larger. For the case of M-1,14,14 foam, the effect is further enhanced by the 
faster growth rate and
higher polydispersivity of this sample.

\begin{figure}
\begin{center}
\includegraphics[width=1\textwidth]{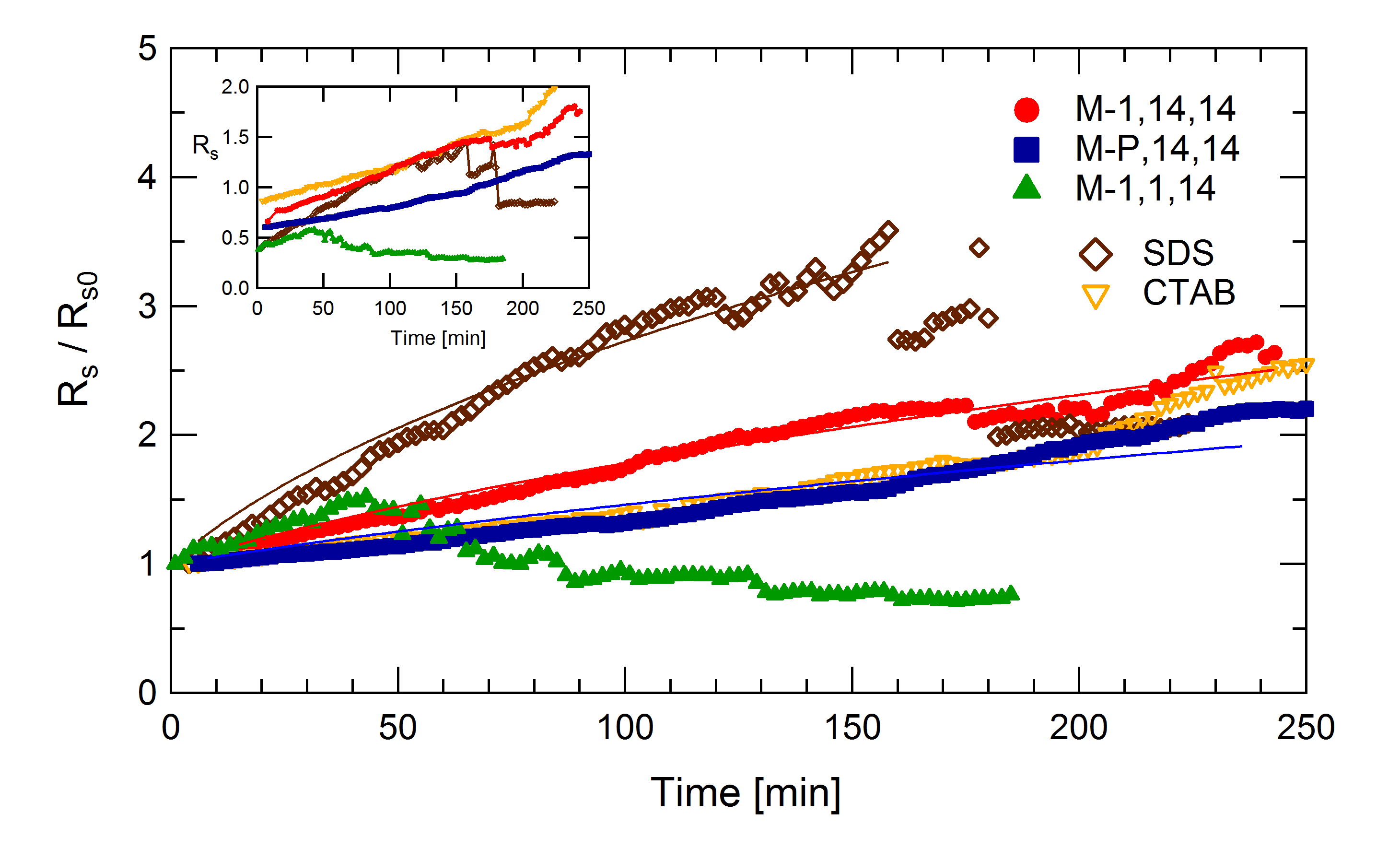}
\caption{Normalized bubble radius \textit{vs.} time for various foams.
The lines are fits of M-1,14,14 (red circles), M-P,14,14 (blue squares) and SDS (brown diamonds) foams to the coarsening model (Eq. 1). The inset shows the time evolution of the average bubble radius $R_s$ of the various samples.
}
\label{fig:bubble_radius}
\end{center}
\end{figure}

We track the average bubble growth of our samples by computing the normalized average bubble radius, $R_s/R_{s0}$,
where $R_{s0}$ is the initial average radius, 
as a function of time. 
The result is 
shown in Fig.~\ref{fig:bubble_radius}. The first thing we observe is that the bubble growth rate of M-1,14,14, M-P,14,14 and CTAB foams are relatively close, while the coarsening rate of SDS foam is much higher by comparison. Note also that the growth rate for SDS halts suddenly at approximately 160 minutes. This happens due to rupture of large bubbles in the top layers. Finally, the coarsening rate of the M-1,1,14 foam initially follows that of SDS before it starts to decrease at about 40 minutes due to severe foam collapse. Subsequent rupture of the larger bubbles in the top layers leads the ratio to dive bellow 1 after 60 min.
The well-known scaling behavior of dry foams is captured by the classical model of foam coarsening \cite{Durian1991Scaling},
\begin{equation}
(R^2(t)-R_0^2)/R_0^2=t/t_c
\end{equation}
where 
$R_0$ is the initial average radius 
and $t_c$ is the coarsening time. As shown in Fig. \ref{fig:bubble_radius}, fits of the coarsening data ($R_s/R_{s0}$) to the model are close in spite of the relative wetness of the foam samples, particularly at earlier times. This is further confirmation that gas diffusion is the chief mechanism of bubble growth for this system, not bubble coalescence.

\begin{figure}
\begin{center}
\includegraphics[width=1\textwidth]{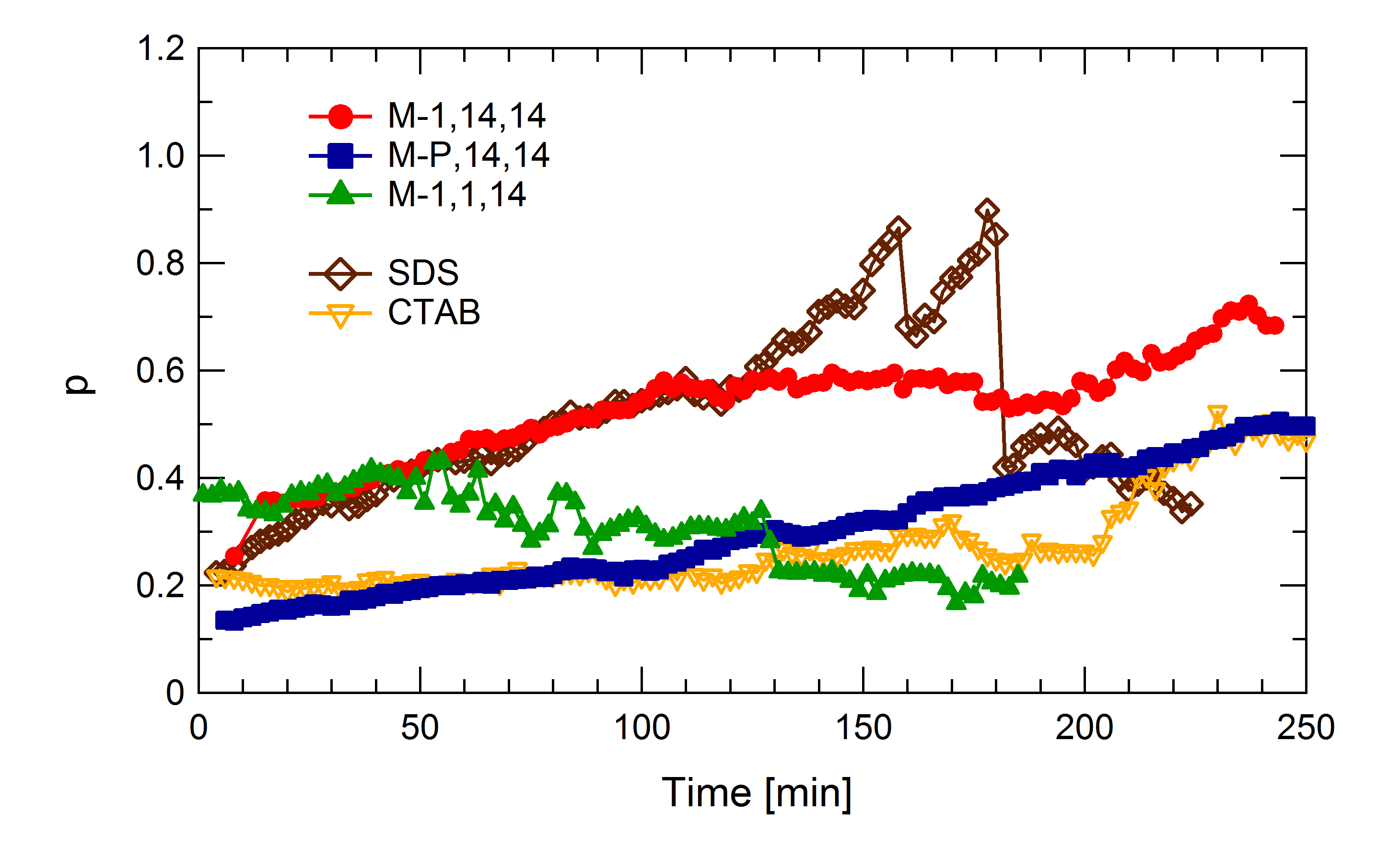}
\caption{Polydispersivity parameter as a function of time for the various foam samples.}
\label{Fig:Polydispersivity}
\end{center}
\end{figure}

One of the outcomes of gas diffusion between bubbles is the broadening of the bubble size distribution \cite{Stavans1993Evolution,Feitosa2006Bubble}. One way to gauge the breadth of the distribution is by computing the polydispersivity parameter ${p=\langle R_s\rangle / \langle R^3 \rangle^{1/3}-1}$~\cite{Kraynik2004Structure}, where $p=0$ corresponds to a monodisperse foam. In Fig. \ref{Fig:Polydispersivity} $p$ is plotted against time for all samples. Initially, M-P,14,14, CTAB and SDS have similar $p$, but the polydispersivity for the SDS foam quickly increase as a result of stronger gas diffusion. Note also that the M-1,14,14 foams starts out more polydisperse, but $p$ increases at a similar pace as the M-P,14,14 and CTAB foams. Lastly, the bubble size distribution of the M-1,1,14 foam starts out broad, but narrows as a result of the rupture of larger bubbles due to instabilities.

\section{Discussion}

Seeking to formulate a coherent picture of the impact of surfactant architecture on foam stability, we start by comparing the molecular structure of the tri-cephalic surfactants with their foamability. Our experiments indicate that: (i) for molecules with the same head group and number of tails (two), those with 14-C tails foam much better than those with 16-C tails; (ii) for molecules with the same head group and tail length (14-C), double-tailed surfactants produce better foam than those with a single tail; (iii) if tails are identical, the identity of the third head group (trimethylammonium or pyridinium) has no effect on foamability, although the surfactant with a trimethylammonium head produces a more polydisperse foam. Analogous results of foamability with respect to the degree of polimerization of DTAB oligomers are reported by Salonen \textit{et al.} \cite{Salonen2010Solutions}.

The investigation of foam stability leads to a similar conclusion: foams made with tri-cephalic double-tailed molecules have better stability than the single-tail one, regardless  of the head structure. Not surprisingly, the coarsening of M-1,14,14 and M-P,14,14 foams (double-tail) is slower than the M-1,1,14 foam (single-tail). For the commercial surfactants, CTAB foam is more stable and has lower coarsening rate, while SDS foam is less table and has higher coarsening rate. This observation matches classical theories to explain foam stability: slow coarsening keeps bubbles small, maintains films wet, and so reduces bubble coalescence. But why is the coarsening slow for these stable foams? 

One possibility is that the initial liquid fraction is higher for the more stable foams. Figure \ref{Fig:liquid_fraction_all}a shows only small differences in the surface liquid fraction profile between the samples. If the differences are taken at face value, one would expect the CTAB and M-1,14,14 foams to possess the highest coarsening rates since they appear to be slightly drier than the others. However, this is not what is observed. Further evidence that the stable foams are initially drier is the value of initial average bubble size. It is well known that foams with small bubbles trap more liquid due to capillary pressure \cite{SaintJalmes2006Physical}. 
In fact, as shown in Table \ref{tab:initial_cond}, the more stable foams have a higher average bubble radius at earlier times and so they should be comparatively drier. 

Coarsening rates in foam can remain low or slow down if the drainage rate slows down. According to the classical result of foam drainage, the velocity of the liquid inside a foam of constant bubble size $R$ is given by \cite{Koehler2000Generalized} 
\begin{equation}
V=K\frac{\rho gR^2}{\mu}\varepsilon^\beta ,
\end{equation}
where $K$ is a dimensionless pre-factor, $\beta$ is an exponent between $1/2$ and 1, $\rho$ is the fluid density, $g$ is the acceleration of gravity, $\mu$ is the fluid viscosity, and $\varepsilon$ is the bulk liquid fraction. The foam permeability $K$ and the exponent $\beta$ depend on the shear viscosity, i.e. on the surface mobility of the surfactant molecules. Most of the drainage in a foam occurs shortly after it is initially produced. During this time, foams made with mobile surfactants should drain faster. As the foam gets drier, drainage slows down and the liquid profile initially set tends to persist,
as illustrated in Fig. \ref{Fig:liquid_fraction_all}b. Consequently,
the surfactant shear viscosity will hardly change the set profiles onwards.  Therefore, it is safe to rule out surface shear viscosity as the prime cause of the enhanced foam stability. A more unusual coupling of drainage and surfactant properties, that goes beyond the classical model, has been proposed by Salonen \textit{et al.}\cite{Salonen2010Solutions} for the case of structurally complex molecules. The idea is that supramolecular structures formed by the aggregation of big amphiphiles jam in the narrow Plateau border channels causing drainage to slow down. Given the size and complex structure of our multi-tail surfactants, it is possible that their micellar aggregates are indeed big. However, it is less certain that this process is slowing down drainage for two reasons. First, the combination of small sample size and capillary pressures keeps the foam relatively wet and the  Plateau borders thick, reducing the possibility of jamming. Second, even if the network channels are thin enough to be jammed, a noticeable difference in coarsening rate should be expected between M-P,14,14 and CTAB foams, since CTAB is known to form small spherical micelles (2-3 nm radius) at the concentrations considered \cite{Goyal1991Shapes,Patel2014PH} and  M-P,14,14 is expected to form larger micelles or even bilayer structures to accommodate the large size of the head group and double tails. While no marked difference in coarsening rate is observed between these two, a major difference is noted with SDS, whose micelle radius is approximately $1.84$ nm \cite{Duplatre1996Size}.  Thus, concerning liquid fraction, it is unlikely that reduced drainage as result of shear viscosity or the entanglement of micelles in the fluid network is the leading cause of low coarsening and enhanced stability of M-1,14,14, M-P,14,14 and CTAB foams.

We investigate two other possible causes for the reduced coarsening rate of the stable foams: the effect of polydispersivity and the  gas permeability of the surfactant monolayer. A foam with a broader distribution of sizes would enhance gas diffusion by larger differences in Laplace pressure \cite{Glazier1987Dynamics}. Hence, the polydispersivity of the foam could be linked to its stability. Looking Fig. \ref{Fig:Polydispersivity} we note that SDS foams show low initial polydispersivity, comparable to the CTAB and M-P,14,14 foams, and yet its coarsening rate is higher and its stability lower. Furthermore, the M-1,14,14 foam, whose polydispersivity better matches that of SDS, has a growth rate closer to its M-P,14,14 counterpart rather than SDS, indicating that polydispersivity does not have a decisive influence on the coarsening rate at this and later stages. Lastly, recent experiments suggest that surfactant structure can have a non-negligible effect on the gas permeation rate through foam films \cite{Farajzadeh2008Foam,Hanwright2005Influence,Tcholakova2011Control}. One of the proposed mechanisms is that long alkyl chains 
pack into condensed states obstructing the transport of gas across the interfaces. The length of the hydrophobic tail, which dictates the solubility of the surfactant, was found to control the gas transport. This can very well be the source of the observed low coarsening rate: bigger, less soluble surfactants are forming less permeable condensed states, while small, more soluble surfactants are forming more permeable fluid states at the interfaces. Whereas liquid fraction and polydispersivity may still contribute in small ways to the stability of the foam, the previous analysis seems to point to the gas permeability of the surfactant monolayer as the chief cause for the slow coarsening rate and ultimately the stability of the foam.

Lastly, it is worth noting that surfactants of good foamability and foam stability have CACs in the range of 0.50 - 1.00 mM. Consulting ref. \cite{Salonen2010Solutions} we note that their easily foamable and stable surfactants also fall in the same range. Furthermore, the combined list indicates that surfactants with a CAC below this range tend to be more difficult to foam, but once they foam, the stability is enhanced. On the other hand, surfactants with a CAC above this range tend to be easy to foam, but also vanish more easily. This suggests that an optimum solubility is necessary to impart strong interfacial properties to the surfactant and combine effortless foam production with durable foams.
So far as we know, Garrett and Moore \cite{Garrett1993Foam} were the first to establish a link between the CAC and optimal foamability and foam stability. Remarkably, their experiment with foams made from alkyl benzene sulfonate blends with carbon chains of various lengths reaches a maximum foamability at intermediate chain lengths for which the CAC is about 0.5 mM. In addition to molecular solubility, maximum foamability is currently thought to be linked to micellar breakdown and subsequent transport to the interfaces \cite{Patist2001Kinetics,Fameau2014Effect}. The observed relationship between foamability and CAC is also related to the hydrophilic-lipophilic balance (HLB) of a surfactant. Determining a value for HLB for these non-conventional ionic surfactants is non-trivial due to their structural complexity and the multiple variables contributing to a meaningful calculation \cite{Pasquali2008Some,James1993Determination,
Laughlin1981HLB}. But in the present study it appears that the M-1,16,16 is too lipophilic, the M-1,1,14 is too hydrophilic and the M-1,14,14 and M-P,14,14 strike the right balance for optimal foamability.

\section{Conclusion}

The advent of new surfactants with a variety of complex structures promises to greatly impact interfacial science and technology \cite{Chevalier2002New}. In many applications, the architecture can actively modify the surface properties and impact the stability of multi-phase systems stabilized by such complex surfactants. Here we have investigated foams made with novel tris-cationic double- and single-tail surfactants. The results show that number of tails and tail length are the critical parameters controlling their foamabilty and foam stability, whereas the nature of the head group seems to have no impact. In the specific case we studied, double-tail surfactants 14 carbons in length easily produce very stable foam as compared to longer double-tail or similar size single-tail surfactants. This result can inform future designs of complex surfactants with the goal of seeking fine control of foam properties.
 
More importantly, when analysed in connection with other physical processes of aging, foam stability was linked to a low coarsening rate of the foam as a result of low gas permeability of the surfactant monolayers rather than other physical processes such as higher liquid fraction or reduced drainage, which are typically attributed as the source of enhanced stability. Therefore, our work demonstrates that for molecules with complex architectures, the gas transport across surfactant monolayers is a parameter that must not be overlooked in investigations of foamability and foam stability. It also encourages more experimental studies to precisely determine how the details of the structure controls the gas transport across the membrane.

Furthermore, our study supports early studies that suggests the critical aggregate concentration as a potentially good predictor of good foamability and stability for foams made with small amphiphilic molecules. Evidently, the collection contemplated in this study is just a very small subset of a vast library of surfactants that include many novel architectures. 
We are particularly interested in expanding the current investigation to a broader range of chain lengths and to observe how robust the relationship is between CAC and foam properties.

\section*{Acknowledgements}

We gratefully acknowledge the financial support provided by the Research Corporation for Science Advancement (KF, SICCSA 21041 and KLC, MICCSA 10709), JMU Center for Material Science, JMU Dept. of Physics and Astronomy, JMU Dept. of Chemistry and Biochemistry, JMU College of Sciecne and Mathematics and the National Science Foundation (CHE - 1062629, CHE - 1461175 and MRI-DMR-1229383). We also thank an anonymous reviewer for pointing us to the work of Garrett and Moore on the link between foam stability and the critical aggregate concentration.

\section*{References}

\bibliography{Klebert}

\begin{thebibliography}{10}
\expandafter\ifx\csname url\endcsname\relax
  \def\url#1{\texttt{#1}}\fi
\expandafter\ifx\csname urlprefix\endcsname\relax\def\urlprefix{URL }\fi
\expandafter\ifx\csname href\endcsname\relax
  \def\href#1#2{#2} \def\path#1{#1}\fi

\bibitem{Menger2000Gemini}
F.~M. Menger, J.~S. Keiper,
  \href{http://www.scopus.com/inward/record.url?eid=2-s2.0-0034595994\&partnerID=40\&md5=8af81303ed214fbfbc82c362cada67c8}{{Gemini
  surfactants}}, Angewandte Chemie - International Edition 39~(11) (2000)
  1906--1920, cited By 742.
\newblock \href
  {http://dx.doi.org/10.1002/1521-3773(20000602)39:11\%3C1906::AID-ANIE1906\%3E3.0.CO;2-Q}
  {\path{doi:10.1002/1521-3773(20000602)39:11\%3C1906::AID-ANIE1906\%3E3.0.CO;2-Q}}.
\newline\urlprefix\url{http://www.scopus.com/inward/record.url?eid=2-s2.0-0034595994\&partnerID=40\&md5=8af81303ed214fbfbc82c362cada67c8}

\bibitem{Shukla2006Cationic}
D.~Shukla, V.~K. Tyagi, \href{http://dx.doi.org/10.5650/jos.55.381}{{Cationic
  Gemini Surfactants: A Review}}, Journal of Oleo Science 55~(8) (2006)
  381--390.
\newblock \href {http://dx.doi.org/10.5650/jos.55.381}
  {\path{doi:10.5650/jos.55.381}}.
\newline\urlprefix\url{http://dx.doi.org/10.5650/jos.55.381}

\bibitem{Marafino2015Colloidal}
J.~N. Marafino, T.~M. Gallagher, J.~Barragan, B.~L. Volkers, J.~E. LaDow,
  K.~Bonifer, G.~Fitzgerald, J.~L. Floyd, K.~McKenna, N.~T. Minahan, B.~Walsh,
  K.~Seifert, K.~L. Caran,
  \href{http://dx.doi.org/10.1016/j.bmc.2015.04.020}{{Colloidal and
  antibacterial properties of novel triple-headed, double-tailed amphiphiles:
  Exploring structure–activity relationships and synergistic mixtures}},
  Bioorganic \& Medicinal Chemistry 23~(13) (2015) 3566--3573.
\newblock \href {http://dx.doi.org/10.1016/j.bmc.2015.04.020}
  {\path{doi:10.1016/j.bmc.2015.04.020}}.
\newline\urlprefix\url{http://dx.doi.org/10.1016/j.bmc.2015.04.020}

\bibitem{LaDow2011Bicephalic}
J.~E. LaDow, D.~C. Warnock, K.~M. Hamill, K.~L. Simmons, R.~W. Davis, C.~R.
  Schwantes, D.~C. Flaherty, J.~A.~L. Willcox, K.~Wilson-Henjum, K.~L. Caran,
  K.~P.~C. Minbiole, K.~Seifert,
  \href{http://dx.doi.org/10.1016/j.ejmech.2011.06.026}{{Bicephalic amphiphile
  architecture affects antibacterial activity}}, European Journal of Medicinal
  Chemistry 46~(9) (2011) 4219--4226.
\newblock \href {http://dx.doi.org/10.1016/j.ejmech.2011.06.026}
  {\path{doi:10.1016/j.ejmech.2011.06.026}}.
\newline\urlprefix\url{http://dx.doi.org/10.1016/j.ejmech.2011.06.026}

\bibitem{Bhattacharya2011Surfactants}
S.~Bhattacharya, S.~K. Samanta,
  \href{http://dx.doi.org/10.1021/jz2001634}{{Surfactants Possessing Multiple
  Polar Heads. A Perspective on their Unique Aggregation Behavior and
  Applications}}, J. Phys. Chem. Lett. 2~(8) (2011) 914--920.
\newblock \href {http://dx.doi.org/10.1021/jz2001634}
  {\path{doi:10.1021/jz2001634}}.
\newline\urlprefix\url{http://dx.doi.org/10.1021/jz2001634}

\bibitem{Haldar2005Synthesis}
J.~Haldar, P.~Kondaiah, S.~Bhattacharya,
  \href{http://dx.doi.org/10.1021/jm049106l}{{Synthesis and Antibacterial
  Properties of Novel Hydrolyzable Cationic Amphiphiles. Incorporation of
  Multiple Head Groups Leads to Impressive Antibacterial Activity}}, J. Med.
  Chem. 48~(11) (2005) 3823--3831.
\newblock \href {http://dx.doi.org/10.1021/jm049106l}
  {\path{doi:10.1021/jm049106l}}.
\newline\urlprefix\url{http://dx.doi.org/10.1021/jm049106l}

\bibitem{Muckom2013Dendritic}
R.~J. Muckom, F.~Stanzione, R.~D. Gandour, A.~K. Sum,
  \href{http://dx.doi.org/10.1021/jp310043a}{{Dendritic Amphiphiles Strongly
  Affect the Biophysical Properties of DPPC Bilayer Membranes}}, J. Phys. Chem.
  B 117~(6) (2013) 1810--1818.
\newblock \href {http://dx.doi.org/10.1021/jp310043a}
  {\path{doi:10.1021/jp310043a}}.
\newline\urlprefix\url{http://dx.doi.org/10.1021/jp310043a}

\bibitem{Maisuria2011Comparing}
B.~B. Maisuria, M.~L. Actis, S.~N. Hardrict, J.~O. Falkinham, M.~F. Cole, R.~L.
  Cihlar, S.~M. Peters, R.~V. Macri, E.~W. Sugandhi, A.~A. Williams, M.~A.
  Poppe, A.~R. Esker, R.~D. Gandour,
  \href{http://dx.doi.org/10.1016/j.bmc.2011.03.036}{{Comparing micellar,
  hemolytic, and antibacterial properties of di- and tricarboxyl dendritic
  amphiphiles}}, Bioorganic \& Medicinal Chemistry 19~(9) (2011) 2918--2926.
\newblock \href {http://dx.doi.org/10.1016/j.bmc.2011.03.036}
  {\path{doi:10.1016/j.bmc.2011.03.036}}.
\newline\urlprefix\url{http://dx.doi.org/10.1016/j.bmc.2011.03.036}

\bibitem{Sugandhi2007Synthesis}
E.~W. Sugandhi, J.~O. Falkinham, R.~D. Gandour,
  \href{http://dx.doi.org/10.1016/j.bmc.2007.03.017}{{Synthesis and
  antimicrobial activity of symmetrical two-tailed dendritic tricarboxylato
  amphiphiles}}, Bioorganic \& Medicinal Chemistry 15~(11) (2007) 3842--3853.
\newblock \href {http://dx.doi.org/10.1016/j.bmc.2007.03.017}
  {\path{doi:10.1016/j.bmc.2007.03.017}}.
\newline\urlprefix\url{http://dx.doi.org/10.1016/j.bmc.2007.03.017}

\bibitem{Paniak2014Antimicrobial}
T.~J. Paniak, M.~C. Jennings, P.~C. Shanahan, M.~D. Joyce, C.~N. Santiago,
  W.~M. Wuest, K.~P.~C. Minbiole,
  \href{http://dx.doi.org/10.1016/j.bmcl.2014.10.018}{{The antimicrobial
  activity of mono-, bis-, tris-, and tetracationic amphiphiles derived from
  simple polyamine platforms}}, Bioorganic \& Medicinal Chemistry Letters
  24~(24) (2014) 5824--5828.
\newblock \href {http://dx.doi.org/10.1016/j.bmcl.2014.10.018}
  {\path{doi:10.1016/j.bmcl.2014.10.018}}.
\newline\urlprefix\url{http://dx.doi.org/10.1016/j.bmcl.2014.10.018}

\bibitem{Weaire2001Physics}
D.~Weaire, S.~Hutzler, \href{http://www.worldcat.org/isbn/0198510977}{{The
  Physics of Foams}}, Oxford University Press, 2001.
\newline\urlprefix\url{http://www.worldcat.org/isbn/0198510977}

\bibitem{Cantat2013Foams}
I.~Cantat, S.~Cohen-Addad, F.~Elias, F.~Graner, R.~Hohler, O.~Pitois,
  F.~Rouyer, A.~Saint-Jalmes, R.~Flatman, S.~Cox,
  \href{http://www.worldcat.org/isbn/0199662894}{{Foams: Structure and
  Dynamics}}, 1st Edition, Oxford University Press, 2013.
\newline\urlprefix\url{http://www.worldcat.org/isbn/0199662894}

\bibitem{2012Foam}
P.~Stevenson (Ed.), \href{http://www.worldcat.org/isbn/0470660805}{{Foam
  Engineering: Fundamentals and Applications}}, 1st Edition, Wiley, 2012.
\newline\urlprefix\url{http://www.worldcat.org/isbn/0470660805}

\bibitem{Langevin2000Influence}
D.~Langevin, \href{http://dx.doi.org/10.1016/s0001-8686(00)00045-2}{{Influence
  of interfacial rheology on foam and emulsion properties}}, Advances in
  Colloid and Interface Science 88~(1-2) (2000) 209--222.
\newblock \href {http://dx.doi.org/10.1016/s0001-8686(00)00045-2}
  {\path{doi:10.1016/s0001-8686(00)00045-2}}.
\newline\urlprefix\url{http://dx.doi.org/10.1016/s0001-8686(00)00045-2}

\bibitem{Addad2013Flow}
S.~C. Addad, R.~H\"{o}hler, O.~Pitois,
  \href{http://dx.doi.org/10.1146/annurev-fluid-011212-140634}{{Flow in Foams
  and Flowing Foams}}, Annual Review of Fluid Mechanics 45~(1) (2013) 241--267.
\newblock \href {http://dx.doi.org/10.1146/annurev-fluid-011212-140634}
  {\path{doi:10.1146/annurev-fluid-011212-140634}}.
\newline\urlprefix\url{http://dx.doi.org/10.1146/annurev-fluid-011212-140634}

\bibitem{Stone2003Perspectives}
H.~A. Stone, S.~A. Koehler, S.~Hilgenfeldt, M.~Durand,
  \href{http://dx.doi.org/10.1088/0953-8984/15/1/338}{{Perspectives on foam
  drainage and the influence of interfacial rheology}}, Journal of Physics:
  Condensed Matter 15~(1) (2003) S283+.
\newblock \href {http://dx.doi.org/10.1088/0953-8984/15/1/338}
  {\path{doi:10.1088/0953-8984/15/1/338}}.
\newline\urlprefix\url{http://dx.doi.org/10.1088/0953-8984/15/1/338}

\bibitem{Tcholakova2011Control}
S.~Tcholakova, Z.~Mitrinova, K.~Golemanov, N.~D. Denkov, M.~Vethamuthu, K.~P.
  Ananthapadmanabhan, \href{http://dx.doi.org/10.1021/la203952p}{{Control of
  Ostwald Ripening by Using Surfactants with High Surface Modulus}}, Langmuir
  27~(24) (2011) 14807--14819.
\newblock \href {http://dx.doi.org/10.1021/la203952p}
  {\path{doi:10.1021/la203952p}}.
\newline\urlprefix\url{http://dx.doi.org/10.1021/la203952p}

\bibitem{Rio2014Unusually}
E.~Rio, W.~Drenckhan, A.~Salonen, D.~Langevin,
  \href{http://dx.doi.org/10.1016/j.cis.2013.10.023}{{Unusually stable liquid
  foams}}, Advances in Colloid and Interface Science 205 (2014) 74--86.
\newblock \href {http://dx.doi.org/10.1016/j.cis.2013.10.023}
  {\path{doi:10.1016/j.cis.2013.10.023}}.
\newline\urlprefix\url{http://dx.doi.org/10.1016/j.cis.2013.10.023}

\bibitem{Hilgenfeldt2001Dynamics}
S.~Hilgenfeldt, S.~A. Koehler, H.~A. Stone,
  \href{http://dx.doi.org/10.1103/physrevlett.86.4704}{{Dynamics of Coarsening
  Foams: Accelerated and Self-Limiting Drainage}}, Physical Review Letters
  86~(20) (2001) 4704--4707.
\newblock \href {http://dx.doi.org/10.1103/physrevlett.86.4704}
  {\path{doi:10.1103/physrevlett.86.4704}}.
\newline\urlprefix\url{http://dx.doi.org/10.1103/physrevlett.86.4704}

\bibitem{Vera2002Enhanced}
M.~U. Vera, D.~J. Durian,
  \href{http://dx.doi.org/10.1103/physrevlett.88.088304}{{Enhanced Drainage and
  Coarsening in Aqueous Foams}}, Physical Review Letters 88~(8).
\newblock \href {http://dx.doi.org/10.1103/physrevlett.88.088304}
  {\path{doi:10.1103/physrevlett.88.088304}}.
\newline\urlprefix\url{http://dx.doi.org/10.1103/physrevlett.88.088304}

\bibitem{Feitosa2008Gas}
K.~Feitosa, D.~Durian, \href{http://dx.doi.org/10.1140/epje/i2007-10329-6}{{Gas
  and liquid transport in steady-state aqueous foam}}, The European Physical
  Journal E: Soft Matter and Biological Physics 26~(3) (2008) 309--316.
\newblock \href {http://dx.doi.org/10.1140/epje/i2007-10329-6}
  {\path{doi:10.1140/epje/i2007-10329-6}}.
\newline\urlprefix\url{http://dx.doi.org/10.1140/epje/i2007-10329-6}

\bibitem{SaintJalmes2006Physical}
A.~Saint-Jalmes, \href{http://dx.doi.org/10.1039/b606780h}{{Physical chemistry
  in foam drainage and coarsening}}, Soft Matter 2~(10) (2006) 836+.
\newblock \href {http://dx.doi.org/10.1039/b606780h}
  {\path{doi:10.1039/b606780h}}.
\newline\urlprefix\url{http://dx.doi.org/10.1039/b606780h}

\bibitem{Langevin2015Bubble}
D.~Langevin, \href{http://dx.doi.org/10.1016/j.cocis.2015.03.005}{{Bubble
  coalescence in pure liquids and in surfactant solutions}}, Current Opinion in
  Colloid \& Interface Science 20~(2) (2015) 92--97.
\newblock \href {http://dx.doi.org/10.1016/j.cocis.2015.03.005}
  {\path{doi:10.1016/j.cocis.2015.03.005}}.
\newline\urlprefix\url{http://dx.doi.org/10.1016/j.cocis.2015.03.005}

\bibitem{Martin2002Network}
A.~Martin, K.~Grolle, M.~Bos, M.~Stuart, T.~Vanvliet,
  \href{http://dx.doi.org/10.1006/jcis.2002.8592}{{Network Forming Properties
  of Various Proteins Adsorbed at the Air/Water Interface in Relation to Foam
  Stability}}, Journal of Colloid and Interface Science 254~(1) (2002)
  175--183.
\newblock \href {http://dx.doi.org/10.1006/jcis.2002.8592}
  {\path{doi:10.1006/jcis.2002.8592}}.
\newline\urlprefix\url{http://dx.doi.org/10.1006/jcis.2002.8592}

\bibitem{Bhattacharyya2000Surface}
A.~Bhattacharyya, F.~Monroy, D.~Langevin, J.-F. Argillier,
  \href{http://dx.doi.org/10.1021/la000320w}{{Surface Rheology and Foam
  Stability of Mixed Surfactant−Polyelectrolyte Solutions{\dag}}}, Langmuir
  16~(23) (2000) 8727--8732.
\newblock \href {http://dx.doi.org/10.1021/la000320w}
  {\path{doi:10.1021/la000320w}}.
\newline\urlprefix\url{http://dx.doi.org/10.1021/la000320w}

\bibitem{Davis2007Comparisons}
J.~P. Davis, E.~A. Foegeding,
  \href{http://dx.doi.org/10.1016/j.colsurfb.2006.10.017}{{Comparisons of the
  foaming and interfacial properties of whey protein isolate and egg white
  proteins}}, Colloids and Surfaces B: Biointerfaces 54~(2) (2007) 200--210.
\newblock \href {http://dx.doi.org/10.1016/j.colsurfb.2006.10.017}
  {\path{doi:10.1016/j.colsurfb.2006.10.017}}.
\newline\urlprefix\url{http://dx.doi.org/10.1016/j.colsurfb.2006.10.017}

\bibitem{Croguennec2006Interfacial}
T.~Croguennec, A.~Renault, S.~Bouhallab, S.~Pezennec,
  \href{http://dx.doi.org/10.1016/j.jcis.2006.06.061}{{Interfacial and foaming
  properties of sulfydryl-modified bovine β-lactoglobulin}}, Journal of
  Colloid and Interface Science 302~(1) (2006) 32--39.
\newblock \href {http://dx.doi.org/10.1016/j.jcis.2006.06.061}
  {\path{doi:10.1016/j.jcis.2006.06.061}}.
\newline\urlprefix\url{http://dx.doi.org/10.1016/j.jcis.2006.06.061}

\bibitem{Salonen2010Solutions}
A.~Salonen, M.~In, J.~Emile, A.~Saint-Jalmes,
  \href{http://dx.doi.org/10.1039/b924410g}{{Solutions of surfactant oligomers:
  a model system for tuning foam stability by the surfactant structure}}, Soft
  Matter 6~(10) (2010) 2271--2281.
\newblock \href {http://dx.doi.org/10.1039/b924410g}
  {\path{doi:10.1039/b924410g}}.
\newline\urlprefix\url{http://dx.doi.org/10.1039/b924410g}

\bibitem{Caran2015Inprep}
T.~M. Gallagher, J.~N. Marafino, B.~K. Wimbish, G.~Fitzgerald, B.~Volkers,
  J.~Floyd, N.~T. Minahan, K.~McKenna, B.~K. Walsh, K.~Thompson, M.~Paneru,
  C.~Dilworth, S.~Djikeng, S.~Masters, S .and~Haji, K.~Seifert, K.~L. Caran,
  Manuscript in preparation.

\bibitem{Rosen2004Surfactants}
M.~J. Rosen, \href{http://dx.doi.org/10.1002/0471670561}{{Surfactants and
  Interfacial Phenomena}}, John Wiley \& Sons, Inc., Hoboken, NJ, USA, 2004.
\newblock \href {http://dx.doi.org/10.1002/0471670561}
  {\path{doi:10.1002/0471670561}}.
\newline\urlprefix\url{http://dx.doi.org/10.1002/0471670561}

\bibitem{Garstecki2004Formation}
P.~Garstecki, I.~Gitlin, W.~DiLuzio, G.~M. Whitesides, E.~Kumacheva, H.~A.
  Stone, \href{http://dx.doi.org/10.1063/1.1796526}{{Formation of monodisperse
  bubbles in a microfluidic flow-focusing device}}, Applied Physics Letters
  85~(13) (2004) 2649--2651.
\newblock \href {http://dx.doi.org/10.1063/1.1796526}
  {\path{doi:10.1063/1.1796526}}.
\newline\urlprefix\url{http://dx.doi.org/10.1063/1.1796526}

\bibitem{Roth2013Structure}
A.~E. Roth, B.~G. Chen, D.~J. Durian,
  \href{http://dx.doi.org/10.1103/physreve.88.062302}{{Structure and coarsening
  at the surface of a dry three-dimensional aqueous foam}}, Physical Review E
  88~(6).
\newblock \href {http://dx.doi.org/10.1103/physreve.88.062302}
  {\path{doi:10.1103/physreve.88.062302}}.
\newline\urlprefix\url{http://dx.doi.org/10.1103/physreve.88.062302}

\bibitem{Feitosa2005Electrical}
K.~Feitosa, S.~Marze, A.~Saint-Jalmes, D.~J. Durian,
  \href{http://dx.doi.org/10.1088/0953-8984/17/41/001}{{Electrical conductivity
  of dispersions: from dry foams to dilute suspensions}}, Journal of Physics:
  Condensed Matter 17~(41) (2005) 6301--6305.
\newblock \href {http://dx.doi.org/10.1088/0953-8984/17/41/001}
  {\path{doi:10.1088/0953-8984/17/41/001}}.
\newline\urlprefix\url{http://dx.doi.org/10.1088/0953-8984/17/41/001}

\bibitem{Durian1991Scaling}
D.~J. Durian, D.~A. Weitz, D.~J. Pine,
  \href{http://dx.doi.org/10.1103/physreva.44.r7902}{{Scaling behavior in
  shaving cream}}, Physical Review A 44~(12) (1991) R7902--R7905.
\newblock \href {http://dx.doi.org/10.1103/physreva.44.r7902}
  {\path{doi:10.1103/physreva.44.r7902}}.
\newline\urlprefix\url{http://dx.doi.org/10.1103/physreva.44.r7902}

\bibitem{Stavans1993Evolution}
J.~Stavans, \href{http://dx.doi.org/10.1088/0034-4885/56/6/002}{{The evolution
  of cellular structures}}, Reports on Progress in Physics 56~(6) (1993) 733+.
\newblock \href {http://dx.doi.org/10.1088/0034-4885/56/6/002}
  {\path{doi:10.1088/0034-4885/56/6/002}}.
\newline\urlprefix\url{http://dx.doi.org/10.1088/0034-4885/56/6/002}

\bibitem{Feitosa2006Bubble}
K.~Feitosa, O.~L. Halt, R.~D. Kamien, D.~J. Durian,
  \href{http://dx.doi.org/10.1209/epl/i2006-10304-5}{{Bubble kinetics in a
  steady-state column of aqueous foam}}, Europhysics Letters 76~(4) (2006)
  683--689.
\newblock \href {http://dx.doi.org/10.1209/epl/i2006-10304-5}
  {\path{doi:10.1209/epl/i2006-10304-5}}.
\newline\urlprefix\url{http://dx.doi.org/10.1209/epl/i2006-10304-5}

\bibitem{Kraynik2004Structure}
A.~M. Kraynik, D.~A. Reinelt, F.~van Swol,
  \href{http://dx.doi.org/10.1103/physrevlett.93.208301}{{Structure of Random
  Foam}}, Physical Review Letters 93~(20) (2004) 208301+.
\newblock \href {http://dx.doi.org/10.1103/physrevlett.93.208301}
  {\path{doi:10.1103/physrevlett.93.208301}}.
\newline\urlprefix\url{http://dx.doi.org/10.1103/physrevlett.93.208301}

\bibitem{Koehler2000Generalized}
S.~A. Koehler, S.~Hilgenfeldt, H.~A. Stone,
  \href{http://dx.doi.org/10.1021/la9913147}{{A Generalized View of Foam
  Drainage:  Experiment and Theory}}, Langmuir 16~(15) (2000) 6327--6341.
\newblock \href {http://dx.doi.org/10.1021/la9913147}
  {\path{doi:10.1021/la9913147}}.
\newline\urlprefix\url{http://dx.doi.org/10.1021/la9913147}

\bibitem{Goyal1991Shapes}
P.~S. Goyal, B.~A. Dasannacharya, V.~K. Kelkar, C.~Manohar, K.~Srinivasa~Rao,
  B.~S. Valaulikar,
  \href{http://dx.doi.org/10.1016/0921-4526(91)90606-f}{{Shapes and sizes of
  micelles in CTAB solutions}}, Physica B: Condensed Matter 174~(1-4) (1991)
  196--199.
\newblock \href {http://dx.doi.org/10.1016/0921-4526(91)90606-f}
  {\path{doi:10.1016/0921-4526(91)90606-f}}.
\newline\urlprefix\url{http://dx.doi.org/10.1016/0921-4526(91)90606-f}

\bibitem{Patel2014PH}
V.~Patel, N.~Dharaiya, D.~Ray, V.~K. Aswal, P.~Bahadur,
  \href{http://dx.doi.org/10.1016/j.colsurfa.2014.04.025}{{pH controlled
  size/shape in CTAB micelles with solubilized polar additives: A viscometry,
  scattering and spectral evaluation}}, Colloids and Surfaces A:
  Physicochemical and Engineering Aspects 455 (2014) 67--75.
\newblock \href {http://dx.doi.org/10.1016/j.colsurfa.2014.04.025}
  {\path{doi:10.1016/j.colsurfa.2014.04.025}}.
\newline\urlprefix\url{http://dx.doi.org/10.1016/j.colsurfa.2014.04.025}

\bibitem{Duplatre1996Size}
G.~Dupl\^{a}tre, M.~F. Ferreira~Marques, M.~da~Gra\c{c}a~Miguel,
  \href{http://dx.doi.org/10.1021/jp960644m}{{Size of Sodium Dodecyl Sulfate
  Micelles in Aqueous Solutions as Studied by Positron Annihilation Lifetime
  Spectroscopy}}, J. Phys. Chem. 100~(41) (1996) 16608--16612.
\newblock \href {http://dx.doi.org/10.1021/jp960644m}
  {\path{doi:10.1021/jp960644m}}.
\newline\urlprefix\url{http://dx.doi.org/10.1021/jp960644m}

\bibitem{Glazier1987Dynamics}
J.~A. Glazier, S.~P. Gross, J.~Stavans,
  \href{http://dx.doi.org/10.1103/physreva.36.306}{{Dynamics of two-dimensional
  soap froths}}, Physical Review A 36~(1) (1987) 306--312.
\newblock \href {http://dx.doi.org/10.1103/physreva.36.306}
  {\path{doi:10.1103/physreva.36.306}}.
\newline\urlprefix\url{http://dx.doi.org/10.1103/physreva.36.306}

\bibitem{Farajzadeh2008Foam}
R.~Farajzadeh, R.~Krastev, P.~L.~J. Zitha,
  \href{http://dx.doi.org/10.1016/j.cis.2007.08.002}{{Foam film permeability:
  Theory and experiment}}, Advances in Colloid and Interface Science 137~(1)
  (2008) 27--44.
\newblock \href {http://dx.doi.org/10.1016/j.cis.2007.08.002}
  {\path{doi:10.1016/j.cis.2007.08.002}}.
\newline\urlprefix\url{http://dx.doi.org/10.1016/j.cis.2007.08.002}

\bibitem{Hanwright2005Influence}
J.~Hanwright, J.~Zhou, G.~M. Evans, K.~P. Galvin,
  \href{http://dx.doi.org/10.1021/la0502894}{{Influence of Surfactant on Gas
  Bubble Stability}}, Langmuir 21~(11) (2005) 4912--4920.
\newblock \href {http://dx.doi.org/10.1021/la0502894}
  {\path{doi:10.1021/la0502894}}.
\newline\urlprefix\url{http://dx.doi.org/10.1021/la0502894}

\bibitem{Garrett1993Foam}
P.~R. Garrett, P.~R. Moore,
  \href{http://dx.doi.org/10.1006/jcis.1993.1315}{{Foam and Dynamic Surface
  Properties of Micellar Alkyl Benzene Sulphonates}}, Journal of Colloid and
  Interface Science 159~(1) (1993) 214--225.
\newblock \href {http://dx.doi.org/10.1006/jcis.1993.1315}
  {\path{doi:10.1006/jcis.1993.1315}}.
\newline\urlprefix\url{http://dx.doi.org/10.1006/jcis.1993.1315}

\bibitem{Patist2001Kinetics}
A.~Patist, S.~G. Oh, R.~Leung, D.~O. Shah,
  \href{http://dx.doi.org/10.1016/s0927-7757(00)00610-5}{{Kinetics of
  micellization: its significance to technological processes}}, Colloids and
  Surfaces A: Physicochemical and Engineering Aspects 176~(1) (2001) 3--16.
\newblock \href {http://dx.doi.org/10.1016/s0927-7757(00)00610-5}
  {\path{doi:10.1016/s0927-7757(00)00610-5}}.
\newline\urlprefix\url{http://dx.doi.org/10.1016/s0927-7757(00)00610-5}

\bibitem{Fameau2014Effect}
A.-L. Fameau, A.~Salonen,
  \href{http://dx.doi.org/10.1016/j.crhy.2014.09.009}{{Effect of particles and
  aggregated structures on the foam stability and aging}}, Comptes Rendus
  Physique 15~(8-9) (2014) 748--760.
\newblock \href {http://dx.doi.org/10.1016/j.crhy.2014.09.009}
  {\path{doi:10.1016/j.crhy.2014.09.009}}.
\newline\urlprefix\url{http://dx.doi.org/10.1016/j.crhy.2014.09.009}

\bibitem{Pasquali2008Some}
R.~C. Pasquali, M.~P. Taurozzi, C.~Bregni,
  \href{http://dx.doi.org/10.1016/j.ijpharm.2007.12.034}{{Some considerations
  about the hydrophilic–lipophilic balance system}}, International Journal of
  Pharmaceutics 356~(1-2) (2008) 44--51.
\newblock \href {http://dx.doi.org/10.1016/j.ijpharm.2007.12.034}
  {\path{doi:10.1016/j.ijpharm.2007.12.034}}.
\newline\urlprefix\url{http://dx.doi.org/10.1016/j.ijpharm.2007.12.034}

\bibitem{James1993Determination}
A.~D. James, J.~M. Wates, E.~Wyn-Jones,
  \href{http://dx.doi.org/10.1006/jcis.1993.1379}{{Determination of the
  Hydrophilicities of Nitrogen-Based Surfactants by Measurement of Partition
  Coefficients between Heptane and Water}}, Journal of Colloid and Interface
  Science 160~(1) (1993) 158--165.
\newblock \href {http://dx.doi.org/10.1006/jcis.1993.1379}
  {\path{doi:10.1006/jcis.1993.1379}}.
\newline\urlprefix\url{http://dx.doi.org/10.1006/jcis.1993.1379}

\bibitem{Laughlin1981HLB}
R.~G. Laughlin,
  \href{http://journal.scconline.org/abstracts/cc1981/cc032n06/p00371-p00392.html}{{HLB,
  from a thermodynamic perspective}}, Journal of Cosmetic Science 32~(6).
\newline\urlprefix\url{http://journal.scconline.org/abstracts/cc1981/cc032n06/p00371-p00392.html}

\bibitem{Chevalier2002New}
Y.~Chevalier, \href{http://dx.doi.org/10.1016/s1359-0294(02)00006-7}{{New
  surfactants: new chemical functions and molecular architectures}}, Current
  Opinion in Colloid \& Interface Science 7~(1-2) (2002) 3--11.
\newblock \href {http://dx.doi.org/10.1016/s1359-0294(02)00006-7}
  {\path{doi:10.1016/s1359-0294(02)00006-7}}.
\newline\urlprefix\url{http://dx.doi.org/10.1016/s1359-0294(02)00006-7}

\end{thebibliography}

\end{document}